\documentclass[sigconf,authorversion,nonacm]{acmart}
\AtBeginDocument{%
  \providecommand\BibTeX{{%
    \normalfont B\kern-0.5em{\scshape i\kern-0.25em b}\kern-0.8em\TeX}}}

\usepackage{graphicx}
\usepackage{booktabs}
\usepackage{todonotes}
\usepackage[section]{placeins}
\usepackage{comment}
\usepackage{subfig}
\usepackage{enumitem}
\usepackage{multirow}
\usepackage{makecell}

\begin{document}
\title[]{Query Smarter, Trust Better? Exploring Search Behaviours\\ for Verifying News Accuracy}
%\title{From Queries and Clicks to Conversations and Blends: How Users Engage with Web and Chat Systems}
\author{David Elsweiler}
\affiliation{
  \institution{University of Regensburg}
  \city{Regensburg}
  \country{Germany}}
\email{david.elsweiler@ur.de}
\author{Samy Ateia}
\affiliation{
  \institution{University of Regensburg}
  \city{Regensburg}
  \country{Germany}}
\email{samy.ateia@ur.de}
\author{Markus Bink}
\affiliation{
  \institution{University of Regensburg}
  \city{Regensburg}
  \country{Germany}}
\email{markus.bink@ur.de}
\author{Gregor Donabauer}
\affiliation{
  \institution{University of Regensburg}
  \city{Regensburg}
  \country{Germany}}
\email{gregor.donabauer@ur.de}
\author{Marcos Fernández Pichel}
\affiliation{
  \institution{University of Santiago de Compostela}
  \city{Santiago de Compostela}
  \country{Spain}}
\email{marcosfernandez.pichel@usc.es}
\author{Alexander Frummet}
\affiliation{
  \institution{University of Regensburg}
  \city{Regensburg}
  \country{Germany}}
\email{alexander.frummet@ur.de}
\author{Udo Kruschwitz}
\affiliation{
  \institution{University of Regensburg}
  \city{Regensburg}
  \country{Germany}}
\email{udo.kruschwitz@ur.de}
\author{David Losada}
\affiliation{
  \institution{University of Santiago de Compostela}
  \city{Santiago de Compostela}
  \country{Spain}}
\email{david.losada@usc.es}
\author{Bernd Ludwig}
\affiliation{
  \institution{University of Regensburg}
  \city{Regensburg}
  \country{Germany}}
\email{bernd.ludwig@ur.de}
\author{Selina Meyer}
\affiliation{
  \institution{University of Regensburg}
  \city{Regensburg}
  \country{Germany}}
\email{selina.meyer@ur.de}
\author{Noel Pascual Presa}
\affiliation{
  \institution{University of Santiago de Compostela}
  \city{Santiago de Compostela}
  \country{Spain}}
\email{noel.pascual.presa@usc.es}

\begin{comment}

\author{David Elsweiler}
\affiliation{
  \institution{University of Regensburg}
  \city{Regensburg}
  \country{Germany}}
\email{david.elsweiler@ur.de}

\renewcommand{\shortauthors}{Mayerhofer, Capra \& Elsweiler}

\end{comment}

\begin{abstract}
While it is often assumed that searching for information to evaluate misinformation will help identify false claims, recent work suggests that search behaviours can instead reinforce belief in misleading news, particularly when users generate queries using vocabulary from the source articles. Our research explores how different query generation strategies affect news verification and whether the way people search influences the accuracy of their information evaluation. A mixed-methods approach was used, consisting of three parts: (1) an analysis of existing data to understand how search behaviour influences trust in fake news, (2) a simulation of query generation strategies using a Large Language Model (LLM) to assess the impact of different query formulations on search result quality, and (3) a user study to examine how 'Boost' interventions in interface design can guide users to adopt more effective query strategies. The results show that search behaviour significantly affects trust in news, with successful searches involving multiple queries and yielding higher-quality results. Queries inspired by different parts of a news article produced search results of varying quality, and weak initial queries improved when reformulated using full SERP information. Although 'Boost' interventions had limited impact, the study suggests that interface design encouraging users to thoroughly review search results can enhance query formulation. This study highlights the importance of query strategies in evaluating news and proposes that interface design can play a key role in promoting more effective search practices, serving as one component of a broader set of interventions to combat misinformation.

\end{abstract}

\begin{comment}

\begin{CCSXML}
<ccs2012>
   <concept>
       <concept_id>10003120.10003121.10003122.10003334</concept_id>
       <concept_desc>Human-centered computing~User studies</concept_desc>
       <concept_significance>500</concept_significance>
       </concept>
   <concept>
       <concept_id>10002951.10003317.10003331.10003336</concept_id>
       <concept_desc>Information systems~Search interfaces</concept_desc>
       <concept_significance>500</concept_significance>
       </concept>
 </ccs2012>
\end{CCSXML}

\ccsdesc[500]{Human-centered computing~User studies}
\ccsdesc[500]{Information systems~Search interfaces}
\end{comment}


\maketitle

\section{Introduction}\label{sec:introduction}

The proliferation of false and misleading information through online media and social networks is a complex, global phenomenon influenced by large online platforms and individual and collective behaviours \cite{lazer2018science, lewandowsky2020technology}. The recent announcement by Meta that it will cease fact-checking \cite{meta2025} highlights a troubling shift in the fight against misinformation. Facebook's efforts had been shown to be effective \cite{allcott2019trends} and studies suggest that fact-checking helps users identify accurate information \cite{schuetz2021combating, zhang2021effects} and boosts trust in social media platforms \cite{acht2024benefitting}. The removal of these efforts just underlines the need to enhance users' information literacy skills. 

While much of the literature focuses on the role of social networks in spreading misinformation, the dominant influence of search engines in shaping the information environment remains under-explored \cite{aslett2024online}. This is particularly significant, since it is widely assumed that searching online to evaluate misinformation would reduce belief in it; this behaviour is common among professional fact-checkers \cite{wineburg2017lateral} and is often taught as part of information literacy interventions \cite{mcgrew2024teaching, breakstone2021lateral}. However, a recent study challenges this assumption. Aslett and colleagues \cite{aslett2024online} report on the results of five large-scale studies demonstrating that online search, when used to evaluate the truthfulness of false news articles, can actually increase the likelihood of believing them. These authors argue that people often use vocabulary from the source article when formulating their queries, which leads to confirmatory results from low-quality information spaces referred to as ``data voids''. They posit that this occurs because specialist terminology is typically not shared between high- and low-quality sources, causing search engines to return corroborating but misleading information.

In this paper, we investigate the impact of various query generation strategies on verifying news articles. We begin by analysing Aslett et al.'s data in greater detail (Section \ref{sec:aslett}), revealing that the way people search plays a larger role in determining trust in a fake news article than the mere act of searching itself. 
Next, we simulate different query generation strategies to examine their impact on search result quality (Section \ref{sec:simulation}). Our findings show that both the parts of the original article used to inspire the initial query and the methods and frequency of query reformulation significantly influence the quality of retrieved results. Finally, we conduct a user study to examine how interface design can help users adopt effective query strategies to improve the accuracy of their information evaluations (Section \ref{sec:user_study}). The findings suggest that while boost strategies had limited success in influencing user behaviour, reading search listings in full appeared to help participants create better queries and achieve improved outcomes, likely by encouraging the use of vocabulary not directly sourced from the article. This suggests several potential avenues for future research that could enhance querying behaviour and, ultimately, more effective news verification.

\section{Related Work}\label{sec:related_work}

We review three relevant bodies of literature: theories and information literacy interventions from the social sciences, contributions from the IR community on misinformation detection, and studies on how people evaluate and trust web-based information, and how this can be positively influenced.

\subsection{Misinformation Research in Social Sciences}

\begin{comment}
In social science, misinformation has been studied from various perspectives, with contributions from fields such as communication, political science, and education \cite{Broda2024, Missau2024}. With some authors arguing that misinformation will always exist and cannot be eliminated from digital spaces, stressing the need for strategies to mitigate its potentially harmful social effects \cite{Wilson2023, Zainudin2024}, the primary focus in this field has been on the detection of misinformation \cite{Sandu2024, Santos2023}.

Various techniques have been shown to have a positive impact on users' capabilities to identify and understand misinformation and unreliable content, such as media literacy strategies \cite{Guess2020} and inoculation interventions \cite{Roozenbeek2022, Traberg2022}. Debunking has also been shown to be an effective strategy if applied precisely. However, since debunking often requires the involvement of other users, its reach can be limited \cite{Helfers2022, Lewandowsky2020}.
Other approaches, such as gamification and "pausing to consider", a technique based on news evaluation, have been shown to reduce the spread of misinformation by improving users' decision making when it comes to sharing such content \cite{Chang2020, Roozenbeek2020, Fazio2020}. 
\end{comment} 

Research on misinformation in the behavioural and social sciences has introduced various means of improving users' competences and behaviours. Disciplines such as cognitive science, political and social psychology, and education research have inspired interventions including debunking false claims \cite{Helfers2022, Lewandowsky2020}, enhancing digital media literacy \cite{Guess2020,badrinathan2021educative}, building resilience against manipulation \cite{Roozenbeek2022, Traberg2022}, slowing the spread of misinformation via the interface design \cite{fazio2020pausing}, subtly encouraging people to think about the accuracy of articles \cite{pennycook2021shifting}, and highlighting the trustworthiness of information \cite{clayton2020real}. The effectiveness of these interventions has been tested using diverse methodologies ranging from controlled experiments (e.g., \cite{wineburg2017lateral,pennycook2021shifting}) to naturalistic field studies (e.g., \cite{Roozenbeek2022,pennycook2021shifting,badrinathan2021educative}). 
Kozyreva et al. compiled a comprehensive toolbox of behavioural and cognitive interventions to combat online misinformation, synthesising evidence from 81 studies conducted worldwide across the social sciences \cite{kozyreva2024toolbox}. The categories of interventions include \textit{nudges}, which subtly influence people's behaviour by altering the environment or context in which decisions are made without restricting their options \cite{thaler2021nudge} as well as \textit{boosts} and educational interventions, which aim to enhance individuals' skills and knowledge, empowering them to make better-informed decisions \cite{Hertwig2017Nudging}.

%The categories of interventions include nudges, which subtly influence people's behaviour by altering the environment or context in which decisions are made without restricting their options e.g. 
%\cite{fazio2020pausing,pennycook2021shifting}; boosts and educational interventions, which aim to enhance individuals' skills and knowledge, empowering them to make better-informed decisions \cite{Guess2020,badrinathan2021educative}; and refutation strategies, such as preemptive inoculation, where people's beliefs are recalibrated by warning them about misinformation before they encounter it \cite{}.

While Large Language Models (LLMs) can hallucinate and amplify the spread of misinformation \cite{shah2024navigating}, they have also been used in personalised conversations to help users reduce their belief in conspiracy theories \cite{costello2024durably, Weeks2023}, showcasing their potential benefits to the field. Given the influence of misinformation, the exploration of new evidence-based approaches to improve users' abilities to recognise such content and limit its spread continues to be a highly relevant research area \cite{Broda2024,Dumitru2022}.

\subsection{Misinformation in IR}

IR has predominantly focused on the detection of %In their comprehensive survey on existing methods to estimate credibility on social media, Viviani and Pasi showed that 
misinformation (see~\cite{viviani2017credibility} for a review). %In the same survey, they also focus on health information on websites, remarking that some features can boost perceived credibility, like the professional appearance of the site. 
%
%Prominent work on misinformation detection includes Castillo et al.~\cite{castillo2011information}, who developed a technique to classify tweets as credible or not credible based on features like the number of re-posts and Sondhi et al., who experimented with different link and content-based features to develop a supervised method to identify unreliable medical webpages~\cite{sondhi2012reliability}, and found that combining all features was the most promising approach. Similarly, Shim et al.~\cite{shim2021link2vec} proposed a model to embed web search results into vectors, which were then fed into traditional ML classifiers, an approach that outperformed conventional fake news detection models. Mazzeo et al. ~\cite{mazzeo2021detection} demonstrated that extracting URL features improved the detection of COVID-19 fake news on web search engines. 
Prominent research includes the work of Castillo et al.~\cite{castillo2011information}, who developed a method to classify tweets as credible or not based on features such as the number of reposts. Sondhi et al.~\cite{sondhi2012reliability} explored link- and content-based features to create a supervised method for identifying unreliable medical webpages, finding that combining all features yielded the best results. Similarly, Shim et al.~\cite{shim2021link2vec} proposed embedding web search results into vectors and using traditional machine learning classifiers, which outperformed standard fake news detection models. Mazzeo et al.~\cite{mazzeo2021detection} showed that extracting URL features enhanced the detection of COVID-19 fake news in web search engines. Recent work has shown that utilising the graph structure of the information ecosystem via graph neural networks (GNNs) can help identifying fake news, e.g. \cite{Dou21User,donabauer2023exploring}.

Initiatives like the \textit{TREC Health Misinformation (HM)} Track, the \textit{CLEF eHealth Consumer Health Search (CHS)} Task, and the \textit{CLEF Check That! Lab} aim to develop retrieval methods that prioritise credible and accurate information over misinformation~\cite{trec20,trec21,suominen2018overview, nakov2021clef}. For example, Pradeep et al.~\cite{pradeep2021vera}, as part of their \textit{TREC HM} participation, proposed a multistage retrieval system with a final supervised re-ranker based on a fine-tuned T5-3b model. In a subsequent study, they integrated LLMs to estimate correct answers to health-related queries and generate query reformulations that improved performance~\cite{pradeep2024towards}. Similarly, Bevendorff et al.~\cite{bevendorff2020webis} used the ChatNoir search engine~\cite{bevendorff2018elastic} for initial candidate retrieval, followed by custom query-based re-ranking.

Events like the \textit{Reducing Online Misinformation through Credible Information Retrieval} workshop (ROMCIR) further advance research in this area~\cite{saracco2021overview,petrocchi2022overview,petrocchi2023romcir,petrocchi2024overview}. These efforts, coupled with increasing interest from the IR community and beyond, underscore that misinformation and its impact on end-users remain unresolved challenges. Addressing this multi-faceted problem extends beyond detecting misinformation, encompassing the study of its effects on searchers, improving information presentation, and empowering users to critically inspect and verify information~\cite{Allan24Future}.

\subsection{(Changing) Search Behaviour}

Research shows that user interaction with search results is often biased \cite{Alonso24Information,BaezaYates18Bias}. One key bias is \emph{position bias}, where users click results based on their placement rather than relevance~\cite{joachims2007evaluating}. Factors like missing or short snippets, absent query terms, and complex URLs further reduce a result's visibility~\cite{clarke2007influence}. Users also tend to favour results that confirm their existing beliefs %,reflecting confirmation bias
~\cite{klayman1987confirmation,nickerson1998confirmation,white2013beliefs,rieger2021item}.

Search result composition influences users' beliefs after searching. Results biased toward incorrect information reduce a user's chance of correctly answering a question, while those favouring correct information improve it~\cite{pogacar2017positive}. Position bias worsens this issue, evidenced by inaccurate featured snippets significantly affecting credibility judgments~\cite{bink2022featured,bink2023investigating}.  %This effect highlights how users anchor their evaluations on the context and presentation of results rather than evaluating them independently as others have identified~\cite{bates1990should,shokouhi2015anchoring}. 
Users are typically inconsistent in judging which search results to trust, with individuals relying on different cues or interpreting the same cue differently~\cite{kattenbeck2019understanding}. This aligns with Prominence-interpretation theory~\cite{fogg2003prominence}, which emphasises the role of subjective perception in trust formation. Users often trust the majority viewpoint in search results, a phenomenon called the \emph{Search Engine Manipulation Effect}~\cite{draws2021sigir}.

To address these issues, search systems have been developed to enhance user decision-making, often drawing on ideas relating to nudge concepts from the social sciences \cite{thaler2021nudge}. For example, result re-ranking algorithms tackle algorithmic biases and improve search quality by rearranging results based on specific criteria~\cite{asudeh2019designing,biega2018equity,celis2017ranking,jaenich2023colbert,zimmerman2019privacy}. Other nudge-like strategies focus on query refinement, encouraging users to explore diverse perspectives or generate alternative ideas~\cite{ahmad2019context,jiang2018rin,niu2014use}. Additionally, some systems provide extra information about results in the snippet to help users make more informed credibility judgments~\cite{schwarz2011augmenting,yamamoto2011enhancing}.

Boost-style interventions have also been applied in search contexts. For instance, search tips enhance user effectiveness~\cite{moraveji2011measuring}, while tools designed to counter confirmation bias encourage users to engage with diverse viewpoints~\cite{rieger2021item,bink2024balancing}. In the area of query generation, interactive examples of high-quality queries during search sessions help users identify key attributes, craft effective queries, and align with expert-level standards~\cite{harvey2015learning}.

This section demonstrates how social science approaches can complement IR and Interactive IR (IIR) research. Here, we integrate methods from these three fields by building on a social science study of search behaviour. First, we conduct a deeper analysis of Aslett et al's data, before applying query simulation techniques commonly used in the IR community to evaluate hypothetical behavioural strategies. Finally, we return to a social science-inspired IIR approach to test whether boost interventions can encourage users to adopt the beneficial behaviours identified in the simulations.

\section{Analysing Aslett et al's data}\label{sec:aslett}

Aslett et al. \cite{aslett2024online} model two conditions in their experiments: with and without search. 
The main goal of their work was to 
study the change on
beliefs about news articles in users who conducted search sessions
to evaluate the truthfulness of the news, and compare this effect with that of users who did not perform any search.
We analyse their publicly available data\footnote{available at \url{https://github.com/SMAPPNYU/Do_Your_Own_Research}} focusing on the search condition, specifically using study 5 data, the only one with query logs. This dataset includes queries, search results, metadata (e.g., NewsGuard scores measuring media quality\footnote{\url{https://www.newsguardtech.com} evaluates the credibility of news websites based on nine criteria, focusing on transparency and journalistic practices, with reviews conducted by trained journalists and editors.}), and participant demographics\footnote{Unfortunately, no click-through or interaction data are available.}. In this first step, we conduct an exploratory analysis to test whether \emph{how} people search is more important than \emph{if} they searched.

The dataset contains 765 queries relating to news articles that were misleading. At the end of the search session the participant made a clear judgement with respect to the veracity of the article. 411 of the queries led to the participant correctly identifying the article as fake news (\emph{Misl}) and 354 (46.3\%) resulted in the participants trusting the content of the article (\emph{True}). 

Queries were slightly longer when participants got it wrong ($\text{mean}_{\text{Misl}} = 6.08$, $\text{sd}_{\text{Misl}} = 4.44$ vs $\text{mean}_{\text{True}} = 6.66$, $\text{sd}_{\text{True}} = 4.23$, $t = -1.8511$, $df = 755.19$, $p$ = 0.06). This was initially surprising since longer queries are typically associated with search experience and expertise~\cite{aula2003query, white2009characterizing,white2007investigating}. 
A second observation that contradicts some prior findings is that successful search sessions —where participants correctly identified fake news— involved more queries on average ($\text{mean}_{\text{Misl}} = 1.82$, $\text{sd}_{\text{Misl}} = 1.21$) than unsuccessful ones, where articles were misclassified as true ($\text{mean}_{\text{True}} = 1.53$, $\text{sd}_{\text{True}} = 0.87$, $t = 3.3283$, $df = 565.38$, $p$ < 0.001). While prior work shows advanced searchers typically issue fewer queries but interact more deeply with results \cite{white2007investigating}, this finding aligns with evidence that domain experts submit more queries than non-experts \cite{white2009characterizing, palotti2016users}.%, suggesting successful participants may have used exploratory strategies.

\begin{figure}[h]
    \centering
    \includegraphics[width=\columnwidth]{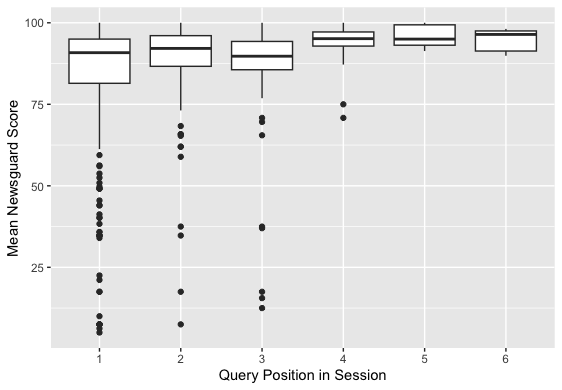}
    \caption{NewsGuard Scores for Search Results against Query Position in the Session}
    \label{fig:newsguard_over_session} 
\end{figure}

Figure \ref{fig:newsguard_over_session} shows one potential reason for the performance improvement when sessions contain more queries. It shows how the average NewsGuard score for the SERP results varies based on the query position in the session. Later queries tend to return results of higher quality ($\rho = 0.15$, $p < 0.001$). 
Moreover, the average NewsGuard scores for search results were higher when participants correctly identified the fake news ($\text{mean}=87.32$, $\text{sd}$=16.41), compared to when they did not ($\text{mean}=82.92$, $\text{sd}$=19.09, $t=3.413, df = 694.8, p< 0.001$). This suggests that the quality of media sources in the search results predicts participants' ability to identify fake news accurately after a search session.

\begin{figure}
    \centering
    \includegraphics[width=\columnwidth]{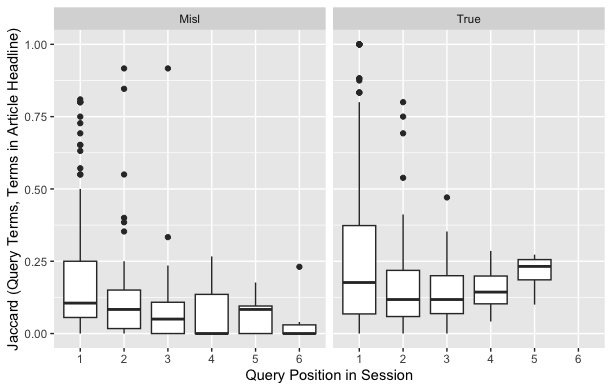}
    \caption{Word Overlap between Query and Article's Headline measured with the Jaccard Coefficient. The left plot represents the cases where users identified the article as false news, while the right plot represents the cases where users believed the article was truthful.} % Replace with your caption
    \label{fig:jaccard_position} % Replace with your label
\end{figure}

\begin{figure}
    \centering
    \includegraphics[width=\columnwidth]{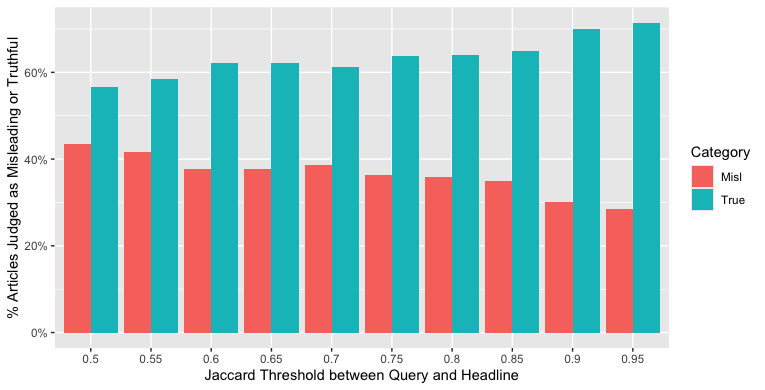}
    \caption{Percentages of  Misleading and Truthful Decisions for Queries with Varying Levels of Overlap with the Article's Headline} % Replace with your caption
    \label{fig:jaccard_overlap} % Replace with your label
\end{figure}

 Aslett et al. suggested that using query terms from the source article leads to confirmatory results. Figures \ref{fig:jaccard_position} and \ref{fig:jaccard_overlap}  confirm that using the article vocabulary is problematic and that the amount of overlap is also important. 
 This explains why longer queries were not always better and why queries sometimes improved as the session progressed. Figure \ref{fig:jaccard_position}
 shows that when participants believed the article was truthful after searching (right plot), the overlap between query terms and the article's headline was high and remained consistent throughout the session. In contrast, when the article was identified as fake news (left plot), the overlap started lower and decreased over consecutive queries. Figure \ref{fig:jaccard_overlap} shows varying levels of overlap between the query and the headline when participants trusted or distrusted the article after searching. This indicates that the percentage of query terms derived from the article's headline is likely a strong predictor for whether participants will classify the article as truthful.

 %Figure \ref{fig:jaccard_overlap}, shows different degrees of overlap between the query and headline when participants trusted or distrusted the article post search. This clearly demonstrates that the percentage of query terms taken from the article's headline is a strong predictor of whether participants will classify the article as truthful. 
 
These initial analyses suggest that the problem with validating news articles does not lie in the act of searching itself but in the way people conduct their searches. The findings indicate that examining querying strategies more closely could provide valuable insights. In the next section, we address this by simulating user query sessions using different querying strategies to assess how these impact the quality of search results, measured by the mean NewsGuard score.

\section{Simulation}\label{sec:simulation}

%Our analysis of Aslett et al.'s data suggests that the source of query terms significantly affects the quality of search results. In this study, we explore this in detail by testing query generation strategies inspired by different sections of the source document. Additionally, we examine how various query reformulation strategies impact performance improvement throughout query sessions. These tested strategies also simulate how users often draw inspiration for queries from search results \cite{}, reflecting known patterns of user interaction with search engines.
Our analysis of Aslett et al.'s data suggests that the source of query terms significantly affects search result quality. To investigate this further, we address two research questions: \textbf{(RQ.S1)} Can we identify effective strategies for generating initial queries by leveraging different sections of the source document? and \textbf{(RQ.S2)} Can we determine effective strategies for reformulating queries to improve performance throughout a search session?

\subsection{Method}

To address these questions, we performed a simulated study that evaluates search performance based on query generation
from different parts of the source article and tests alternative reformulation strategies that simulate how users derive inspiration from search results. This reflects established patterns of user interaction with search engines (see review above). The IR community has a strong tradition of employing simulated query generation studies, which enable systematic, controlled, and efficient testing of query strategies \cite{azzopardi07,Maxwell19Modelling,elsweiler21}, including reformulating queries throughout a session \cite{gunther2022,hagen13,gunther21}, without human users.

After preliminary experiments with traditional and newer simulation approaches, including methods oriented to sample queries from classic language models \cite{carterette15} and neural techniques based on docT5query \cite{nogueiralin2019} and keyBERT \cite{grootendorst2020keybert},
we decided to use a Generative-AI approach powered by Llama3 (see repository for specific technical details)\footnote{Full details of the simulation process, including code, prompts and the queries themselves can be found here: \url{https://anonymous.4open.science/r/search-verify-simulation-7AB1}}. This decision was based 
on the fact that traditional methods tended to produce unrealistic queries, often drifting away from the topic of the source article. The problem of topic drift in query session simulation has been discussed in the literature \cite{gunther21} and, thus, we manually inspected the simulated queries to ensure that topic drift was not an issue in the final configuration of the simulations.

For our simulation process, we started with the 17 articles labelled as fake news in Aslett et al. \cite{aslett2024online} study 5\footnote{Since 9 of these articles were no longer available online at the time of our study, we used the Wayback machine to recover these.}. %Our main goal was to simulate search sessions a user could conduct to verify the information in the original articles. 
We used Bing search API for searching and Llama3 (8B parameters) to generate the synthetic queries. We hypothesised that focusing on different aspects of the initial article and the search results during query generation can impact the quality of the results. To test this, we developed a series of query generation variants. A key feature of these approaches is their interpretability, allowing them to be translated into concrete behavioural strategies that human users could easily implement.

\begin{itemize}
\item \textbf{Initial query generation strategies}: We instructed the LLM to build queries after reading the article headline (\textbf{H} variant), after reading the headline and the first paragraph (\textbf{H 1P}), or after reading the full text of the article (\textbf{FT}). The respective parts of the article were fed to the LLM together with instructions for the target search task (news verification).
The generated query was run against the Bing search API, obtaining the first top 10 results of the search session. 

\item \textbf{Query reformulation strategies}: For the follow-up searches within the session, we tested two reformulation approaches. The first approach instructs the LLM to consider the previous queries in the session and the title and snippets from the top 10 search results of the last search (\textbf{TS} reformulation variant). The second approach also forces the LLM to consider the previous queries, the title \& snippets from the top 10
results but additionally feeds 
the first paragraph of the top two search results (\textbf{TS 1P TOP2} reformulation variant). This approach simulates typical user behaviour (the first two results are much more likely to receive user clicks \cite{joachims2007evaluating})\footnote{We restricted the input to the LLM to the first paragraphs from the top 2 pages because, otherwise, the context becomes too lengthy and noisy. Furthermore, 
these leading paragraphs arguably 
reflect the parts of these pages
that are more likely read by web users.}.
\end{itemize}

In total, we tested six different search variants (three initial query generation * two query reformulation 
strategies). We simulated search sessions of lengths from one to five, based on the lengths in the sessions in Aslett et al's study. 
Each variant was executed ten times to minimise randomness in the results. We instructed the LLM to generate queries of approximately 3 to 5 words, based on the query lengths in the real data. Only search results published on the same day of the article or earlier were considered. All other results were removed as to simulate web pages available online at the date of publication of the article.

As a quality measure, we used the mean NewsGuard score of the search results since we previously demonstrated its correlation with users making better decisions. We observed that the SERPs of the simulated sessions showed good coverage of pages that have a NewsGuard score assigned
(over 50\% of the retrieved webpages, which is higher than the percentage of NewsGuard-scored pages in the original user study, 30\%). This guarantees that the SERPs of the simulation can be assessed with sufficient confidence.

%To assess its feasibility, we conducted test sessions and found that over 50\% of the SERP entries had a NewsGuard score, compared to around 30\% in Aslett et al.'s original data. We suspect this is because lower-quality results are gradually removed from search indexes over time. 

\begin{comment}
The simulation aims to answer the following research questions:

\begin{itemize}
    \item \textbf{RQ.S1}: Can we determine good strategies for generating initial queries?
    \item \textbf{RQ.S2}: Can we determine good strategies for re-formulating queries?
\end{itemize}
\end{comment}

\subsection{Results}

The following subsections present the simulation study results:

\subsubsection{Initial Query Generation}

\begin{comment}
{RQ.S1} examines the best strategy for generating initial queries. Notably, all three tested strategies produced high NewsGuard scores for the SERPs of the first search: the H variant achieved an average score of $\text{mean}_\text{H}=94.58$, $\text{sd}_\text{H}=4.6$, while H 1P and FT scored $\text{mean}_{\text{H1P}}=95.26$, $\text{sd}_{\text{H1P}}=3.9$ and $\text{mean}_{\text{FT}}=94.10$, $\text{sd}_{\text{FT}}=5.2$, respectively. These results are significantly higher than the scores observed in Aslett's data. This aligns with our observation that many fake news articles from the original dataset had disappeared over time, requiring retrieval through the Wayback Machine. We believe this reflects a dual process: fake news articles are often removed from publication (the articles were over two years old by the time of our simulation), and search engines adapt by demoting or removing low-quality content as declining clicks lower their rankings. This likely explains the higher percentage of results with NewsGuard scores in our simulated study.
\end{comment}

{RQ.S1} examines the best strategy for generating initial queries. Notably, all three tested strategies produced high NewsGuard scores for the SERPs of the first search. The H variant achieved an average score of $\text{mean}_\text{H} = 94.58$ (SD = 4.6), while H 1P and FT scored $\text{mean}_{\text{H1P}} = 95.26$ (SD = 3.9) and $\text{mean}_{\text{FT}} = 94.10$ (SD = 5.2), respectively. These results are significantly higher than the scores observed in Aslett's data. %This aligns with our observation that many fake news articles from the original dataset had disappeared over time and had to be retrieved through the Wayback Machine. 
We believe this reflects a dual process: fake news articles are often removed from publication (the articles were over two years old by the time of our simulation), and search engines adapt by demoting or removing low-quality content as declining clicks lower their rankings. This likely also explains the higher percentage of results with NewsGuard scores in our simulated study.

\begin{comment}
We compared the initial query generation strategies 
and found that using the headline and the first paragraph (H 1P) provides search results
whose average
NewsGuard score 
is significantly higher
than those obtained
with the full text (FT) 
strategy ($\text{mean}_{\text{H1P}} = 95.26$, $\text{sd}_{\text{H1P}} = 3.9$ vs $\text{mean}_{\text{FT}} = 94.10$, $\text{sd}_{\text{FT}} = 5.2$, $U = 59355$, $df = 658$, $p = 0.04$)\footnote{We used non-parametric Mann-Withney U test for these comparisons}. Using the full text (FT) was the least effective approach, and relying solely on the headline (H) also produced slightly lower quality SERPs, though the difference with H 1P was not statistically significant ($\text{mean}_{\text{H1P}} = 95.26$, $\text{sd}_{\text{H1P}} = 3.9$ vs $\text{mean}_{\text{H}} = 94.58$, $\text{sd}_{\text{H}} = 4.6$, $U = 48802$, $df = 648$, $p = 0.09$). 
The comparison between FT and H yielded no significance ($\text{mean}_{\text{FT}} = 94.10$, $\text{sd}_{\text{FT}} = 5.2$ vs $\text{mean}_{\text{H}} = 94.58$, $\text{sd}_{\text{H}} = 4.6$, $U = 51587$, $df = 648$, $p = 0.61$). 
From these results, we can derive that the most effective strategy generates queries from both the headline and the first paragraph (H 1P), perhaps meaning that the first paragraph may contain more neutral language, while later sections of the document may include misinformation cues —unique terms or phrases that can lead users toward data voids.
\end{comment}

We compared the initial query generation strategies and found that using the headline and the first paragraph (H 1P) resulted in search results with a significantly higher average NewsGuard score than those obtained with the full text (FT) strategy. Specifically, the mean for H 1P was 95.26 (SD = 3.9), compared to 94.10 (SD = 5.2) for FT, with a U value of 59355, $df = 658$, and $p = 0.04$\footnote{We used a non-parametric Mann-Whitney U test for these comparisons}. The full text (FT) strategy was the least effective, and using only the headline (H) produced slightly lower quality search results. However, the difference between H and H 1P was not statistically significant: the mean for H 1P was 95.26 (SD = 3.9) and for H it was 94.58 (SD = 4.6), with a U value of 48802, $df = 648$, and $p = 0.09$. The comparison between FT and H also showed no significance, with a mean of 94.10 (SD = 5.2) for FT and 94.58 (SD = 4.6) for H, a U value of 51587, $df = 648$, and $p = 0.61$. From these results, we can conclude that the most effective strategy is to generate queries from both the headline and the first paragraph (H 1P), suggesting that the first paragraph may contain more neutral language, while later sections of the document might introduce misinformation cues —unique terms or phrases that can lead users toward data voids.

%We used the Mann Whitney U Test ($\alpha=.05$) to compare the strategies. Using the headline and the first paragraph (H 1P) is significantly better than using the full text (FT) for the initial generation. The rest of pairwise comparisons yielded no significant difference. The findings indicate that counter-intuitively using the full text (FT) is the least effective approach, and relying on the headline (H) alone is also suboptimal as, although not statistically significant difference, it produces slightly lower quality SERPs. The most effective strategy seems combining the headline with the first paragraph (H + 1P). This result might indicate that only reading the headline might not be enough, but also considering the entire document might be counterproductive.

\subsubsection{Query Reformulation Strategies}

\begin{table}
    \centering
    \caption{Average NewsGuard Score in the SERPs (first and fifth query in the session)}
    \begin{tabular}{ccc}
        \toprule
        \textbf{Simulation variant} & \textbf{First query} & \textbf{Fifth reformulation} \\
        \midrule
        H - TS & 94.93 & 95.38 \\
        H - TS 1P TOP2 & 95.00 & 95.65 \\
        H 1P - TS & 94.94 & 95.48 \\
        H 1P - TS 1P TOP2 & 96.07 & 95.79 \\
        FT - TS & 95.00 & 95.61 \\
        FT - TS 1P TOP2 & 94.43 & 95.22 \\ 
        \toprule
    \end{tabular}
    \label{tab:fifth_reform}
\end{table}

{RQ.S2} explores the most effective reformulation strategy for search queries. We compared different strategies by evaluating the average NewsGuard scores from all queries in the simulated sessions.

The comparison between FT-TS and FT-TS 1P TOP2 revealed a significant difference ($U = 307816$, $df = 1644$, $p = 0.001$), with FT-TS having a mean of 94.9 (SD = 5.1) and FT-TS 1P TOP2 a mean of 94.2 (SD = 5.2). No significant differences were found for other pairwise comparisons between reformulation strategies.
\begin{comment}
    
The comparison between H - TS and H - TS 1P TOP2 showed no significant difference. The mean for H - TS was 94.6 (SD = 4.9), and for H - TS 1P TOP2 it was 94.9 (SD = 4.2), with a U value of 298412, $df = 1558$, and $p = 0.52$. Similarly, there was no difference between H 1P - TS and H 1P - TS 1P TOP2. The mean for H 1P - TS was 95.2 (SD = 3.8), and for H 1P - TS 1P TOP2 it was 95.3 (SD = 3.9), with a U value of 314603, $df = 1612$, and $p = 0.24$. However, the comparison between FT-TS and FT - TS 1P TOP2 revealed a significant difference. The mean for FT-TS was 94.9 (SD = 5.1), and for FT - TS 1P TOP2 it was 94.2 (SD = 5.2), with a U value of 307816, $df = 1644$, and $p = 0.001$. This indicates that the TS reformulation strategy was the most effective.
\end{comment}
These results suggest that if the initial query generation strategy is strong (H 1P or H), the specific reformulation strategy has little impact. However, when the initial query is poor (FT), reformulations inspired only by the entire search engine results page (TS) yield significantly better results than those achieved by TS 1P TOP2. This makes sense as a weak query might lead us to poor top results and, thus, if we reformulate the query guided by the top 2 results (TS 1P TOP2) we might be getting to even poorer results. Observe also that the most effortful reformulation strategy 
(TS 1P TOP2, which involves not only inspecting the SERP but also reading a couple of paragraphs from the top 2 results) is not the most effective.

We analysed the evolution of NewsGuard scores during search sessions and found that query reformulation reduces performance differences, with the most variation in the first query. By the fifth reformulation, differences between approaches are minimal (see Table~\ref{tab:fifth_reform}). While this might seem reassuring, Aslett's data show us that web users rarely generate that many queries in a single session.

%We also analysed the evolution of Newsguard scores within the search session and found that query reformulation tends to reduce differences in performance, with the greatest variation observed in the first query. By the fifth reformulation, differences between approaches become minimal, see Table~\ref{tab:fifth_reform}. This might seem reassuring but note that web users are unlikely to generate that many queries in the session.

These simulation findings align with Aslett et al.'s data, emphasising the importance of query reformulation. While most users do not reformulate extensively, effective reformulation becomes critical when the initial query is weak, enabling performance improvements with a few reformulations.

\begin{figure}
    \centering
    \includegraphics[width=\linewidth]{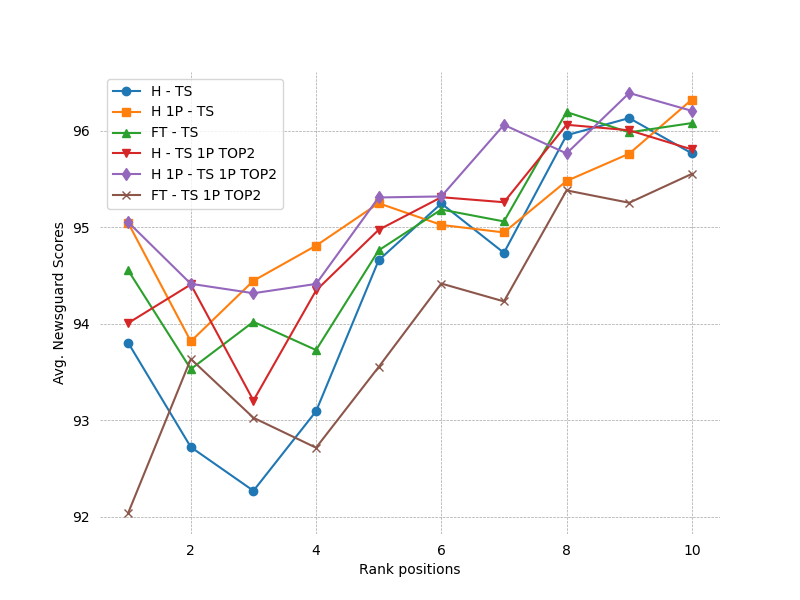}
    \caption{NewsGuard Scores  at different rank positions}
    \label{fig:sim_rank}
\end{figure}

We also analysed how NewsGuard Scores vary at different rank positions. As shown in Figure~\ref{fig:sim_rank}, these scores tended to increase with the ranking positions, at least up to position 10.
This trend occurred for all variants and suggests that the most reputed results (at least according to NewsGuard assessment) are not necessarily at the top positions. 
Web users should, therefore, inspect the full SERP and not only the top positions. This effect could be due to the tendency of search engines to promote popularity (e.g., with link-based metrics), but popularity does not equal reputation. 

This plot also confirms that FT-* methods are inferior to their counterparts. 
H1P-* methods yield the highest proportion of high quality docs at top positions, where users more likely click. Again, we ran statistical tests that confirmed that the 
H1P-* variants were superior for the top entries in the rank. 

\begin{table}[t!]
\centering
\caption{Jaccard overlap between queries from different simulation variants and article headline (H) or first paragraph (1st)}
\label{tab:jaccard_scores}
\begin{tabular}{lcccc}
\toprule
\textbf{Variant} & \parbox[c]{1.1cm}{\centering \textbf{Mean}\\ J(Q,H)} & \parbox[c]{1.1cm}{\centering \textbf{SD}\\ J(Q,H)} & \parbox[c]{1.1cm}{\centering \textbf{Mean}\\ J(Q,1st)} & \parbox[c]{1.1cm}{\centering \textbf{SD}\\ J(Q,1st)} \\
\midrule 
FT TS & 0.12 & 0.11 & 0.04 & 0.04 \\
FT TS 1P TOP 2 & 0.10 & 0.10 & 0.05 & 0.04 \\
H 1P TS & 0.10 & 0.08 & 0.04 & 0.04 \\
H 1P TS 1P TOP 2 & 0.08 & 0.08 & 0.03 & 0.04 \\
HTS & 0.09 & 0.08 & 0.03 & 0.03 \\
HTS 1P TOP 2 & 0.09 & 0.09 & 0.02 & 0.03 \\
\toprule
\end{tabular}
\end{table}

Given our findings in Section \ref{sec:aslett}, it was surprising that the headline conditions performed the best. To better understand why this was the case, we examined the Jaccard overlap between the generated queries and both the source article's headline and first paragraph (see Table \ref{tab:jaccard_scores}). The results reveal that the best-performing variants had, on average, the least overlap with the source article. In other words, supplying the model with text does not guarantee that the model will merely extract words from that text to generate the query. Indeed, when more of the source article was provided, the query terms were more likely to be sourced from there. This may also apply to human users.
%[DE: I think this might be a good place to put the analyses of the simulated queries since there is a conflict between the impression given here- query terms from the headline are good- and those in Sections 3 and 5 - query terms from the headline are bad. The analysis shows that despite having access to the headline, the LLM did not use it (for whatever reason)]

%Additionally, a positive takeaway is that Newsguard scores are, on average, very high. As shown in Figure x, 

\section{Empowering Users to Query Better}\label{sec:user_study}

In this section we perform a pre-registered user study\footnote{\url{https://osf.io/x9g74/?view_only=c1cef259191c4a8dabd0602b1a2c1470}} to investigate how boost interventions can empower users to act more effectively based on the search tactics discussed earlier. Based on the results from our previous analyses, we draw four key conclusions, with the mapping to user study conditions provided in italics:

\begin{enumerate}
    \item Initial queries were more effective when the algorithm had access to the headline and first paragraph, but not the rest of the article \textit{(Read 1st)}.
    \item Strong queries tend to have less overlap with the text of the source article \textit{(Own Words)}.
    \item Weak initial queries improve by reviewing the entire results page, which is more likely to yield reliable documents than focusing solely on the top results \textit{(Read All)}.
    \item Submitting more queries within a session improves result quality by the fifth query, regardless of the reformulation strategy used \textit{(Multiple Queries)}.
\end{enumerate}

\begin{comment}
Our analysis in Section \ref{sec:aslett} highlights the impact of search tactics on result quality and belief in fake news. The simulation results in Section \ref{sec:simulation} confirm that the information available prior to the initial generation and reformulation phases is crucial. From the results, we draw four key conclusions (the mapping to user study conditions is given in italics): 1. Initial queries were more effective when the algorithm had access to the headline and first paragraph and not the rest of the article \textit{(Read 1st)}. 2. Strong querires tend to have lower overlap with the text of the source article \textit{(Own Words)}. 3. Weak initial queries can be improved by inspecting the entire results page, which is more likely to lead to reliable documents than focusing solely on the top search results \textit{(Read All)}.  4. Submitting more queries within a session improves result quality by the fifth query, regardless of the reformulation strategy used \textit{(Mult. Queries)}. Here, we examine by means of a pre-registered user study\footnote{\url{https://osf.io/x9g74/?view_only=c1cef259191c4a8dabd0602b1a2c1470}} to what extent we can empower users to act effectively in these situations using boost interventions.
\end{comment}

Inspired by the literature, we developed boost messages (one for each condition) that aim to encourage the successful behaviours observed in our previous analyses (see Table \ref{tab:study_conditions} for an overview and Figure \ref{fig:interface} for the boost presentation in the SERP). 

\begin{table}
    \centering
    \caption{Study Conditions with either a vanilla SERP or a boost containing a search tip.}
    \begin{tabular}{p{0.25\linewidth}p{0.65\linewidth}}
        \toprule
         \textbf{Boost}& \textbf{Wording}\\
        \midrule 
         No boost& Vanilla SERP w/o boost\\
         \midrule 
 Own Words&Users in our pre-study, who formulated their queries in their own words —rather than simply extracting keywords from the source document— were far more successful at detecting fake news.\\
 \midrule 
 Read 1st&Users in our pre-study who \emph{fully} read the first paragraph of an article before formulating their queries were far more successful at detecting fake news.\\
 \midrule 
 Read All&Users in our pre-study who \emph{read all search results} before crafting their query were far more successful at detecting fake news\\
 \midrule 
 Multiple Queries&Users in our pre-study who issued the most queries were the most successful at detecting fake news.\\
 \toprule
    \end{tabular}
    \label{tab:study_conditions}
\end{table}

\subsection{Materials and Setup}

Participants performed the same task as in Aslett et al., searching to assess the trustworthiness of news articles. Participants were assigned to only one condition (between groups) and evaluated only a single article. All articles used in this study were identified as misleading, aligning with the prior analyses. Articles were selected from outlets included in the Aslett et al. dataset that provided articles their fact-checkers classified as misleading. Both conservative and liberal sources were represented, including the following websites: \textit{The Federalist Papers}, \textit{ZeroHedge}, \textit{Natural News}, \textit{WND}, \textit{Stillness in the Storm}, \textit{Palmer Report}, \textit{GNews}, \textit{Occupy Democrats}, \textit{Townhall}, and \textit{Newsmax}.

To identify suitable articles, these websites were scraped, and URLs of current articles were collected. A predefined prompt was used with GPT-4 as a tool to assist in the evaluation process. A team of three researchers manually examined fresh articles (no more than two days old) and identified candidates they believed were fake news. The prompt provided guidance by offering an automated likelihood score based on source credibility, content analysis (bias, sensationalism, unsupported claims), and cross-verification with credible sources. Articles flagged by the researchers as potentially fake news were then sent to professional fact-checkers for verification, ensuring that only articles verified to be misleading were included in the study\footnote{The prompt and fact-checker reports on the articles are included in our repository: \url{https://anonymous.4open.science/r/sigir-aslett-misinfo-BC03}}.

%In this study, aligning with our analyses above, all articles are fake news,  selected from outlets included in the Aslett et al. dataset that provided untrustworthy news and verified as fake by professional fact-checkers.\footnote{Full details of the selection of the articles, the fact checking process and access to the data will be provided in an Appendix} 

\begin{figure}
    \centering
    \includegraphics[width=\linewidth]{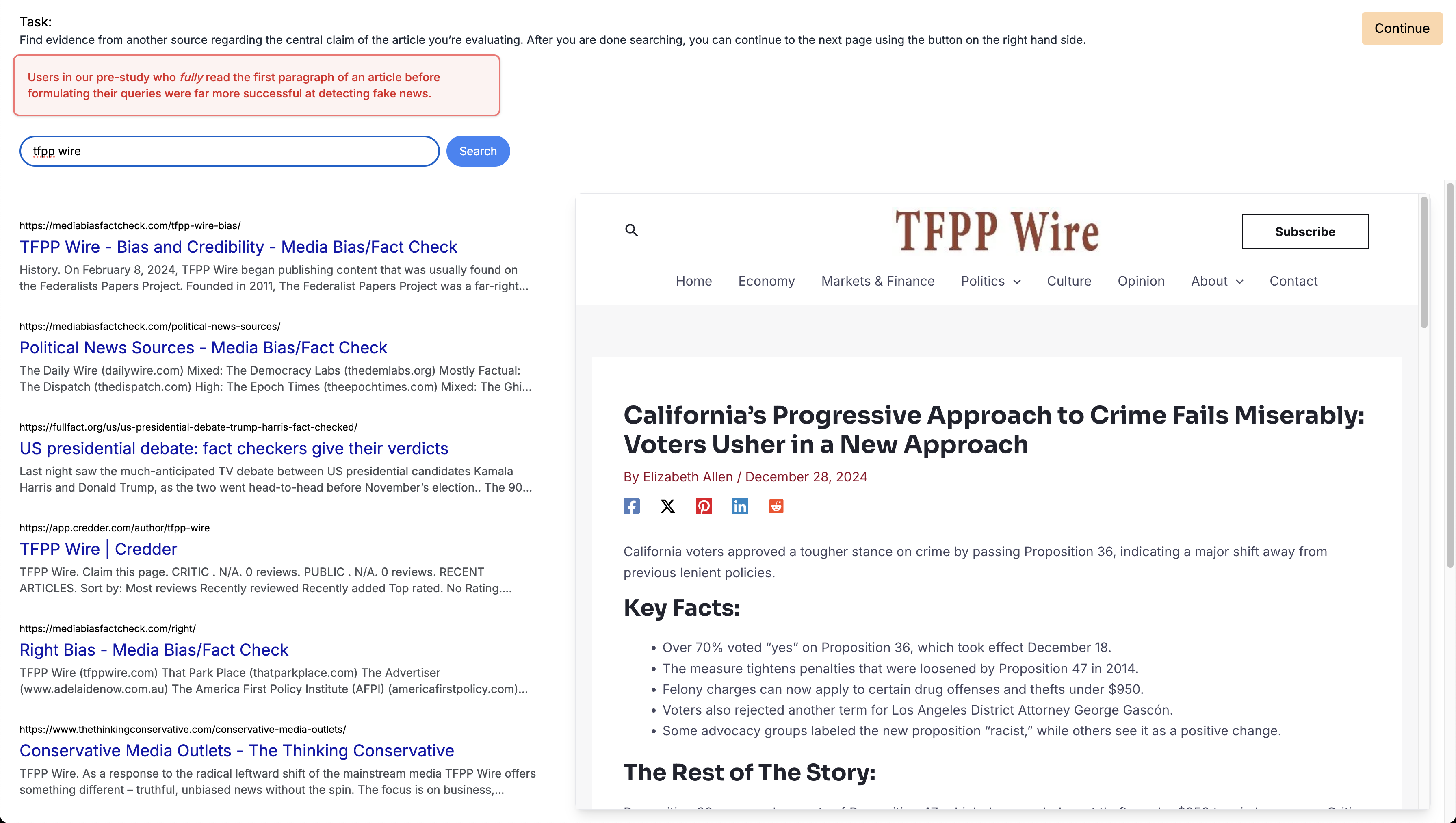}
    \caption{The search interface used in our study, boost shown in red in the upper left-hand corner}
    \label{fig:interface}
\end{figure}

By using fresh articles, we assume that lower-quality results remain in the search index, addressing an issue observed in the simulated study. We attained four articles from four different media outlets. Searches were conducted through an experimental system (see Figure \ref{fig:interface}) powered by the Bing API, enabling us to record detailed interaction data including queries submitted, result clicks, and timestamps.    
Participants were provided with the news article in html form on the right-hand pane and could search using the interface on the left side. The boost was provided in a prominent position in the top-left of the screen to make it more likely to be read. As in Aslett et al \cite{aslett2024online}, after searching participants evaluated the article and provided demographic information and self-estimated digital literacy and veracity scores.\footnote{See Appendix J in Aslett et al's paper \cite{aslett2024online} for the questions}

\subsection{Hypothesis}

We define the following hypothesis %: 
%\begin{itemize}
%    \item 
    H1: \textit{Users in the boost conditions will submit queries resulting in higher average NewsGuard scores}.
    % I removed the additional hypotheses since these complicate matters. I think these are better suited to an exploratory analysis anyway.
    %\item H2: Users in the boost condition will submit more queries on average over the search sessions
    %\item H3: Users in the boost condition will have higher average NewsGuard scores for 2nd and 3rd queries submitted in a search session.
%\end{itemize}

We focused on NewsGuard scores because we assumed they are less influenced by user factors, such as political views and education, or article characteristics, such as presentation quality. Additionally, Aslett's data show that mean NewsGuard scores strongly predict search outcomes when these other variables are controlled.

\subsection{Participants}

The number of participants was established by means of a power analysis. 
Given that the number of queries submitted will vary per participant, we initially considered a mixed-effects model. However, due to the low number of repeated measures for many participants (40\% submitted only 1 query), attempts to fit a random-effects model based on the data simulated for the power analysis resulted in singular fits. Consequently, we opted for a fixed-effects model (one-way ANOVA) to compare the five experimental conditions. A power analysis determined that 200 participants are required to achieve 80\% power at an alpha level of 0.05, assuming a medium effect size (\textit{f} = 0.25). Full details of the power analysis can be found in the pre-registration.

A total of 260 participants were recruited via Prolific to obtain 200 participants (120 male, 77 female, and 3 diverse/other) who passed the attention check. These were all US-based native or fluent English speakers, 106 of whom hold a bachelor's degree, 31 a master's, 49 a high school diploma, and 9 a doctorate. Ages ranged from 19 to 72 years, with a median of 36 and a mean of 37.17 (IQR: 26.75--45). Participants reported a wide range of occupations, including roles in IT (e.g., software engineers, data analysts), healthcare (e.g., nurses, physicians), education (e.g., teachers, researchers), business (e.g., project managers, administrators), with 11 participants identifying as students and 7 as unemployed.

On average, participants took 7.09 minutes to complete the study ($SD = 4.73, Mdn = 5.68)$. This varied slightly based on the condition with those in the 'Multiple Queries' condition ($M = 8.72, SD = 6.81, Mdn = 6.00$) taking the longest to complete and 'Read 1st' ($M = 6.21, SD = 3.93, Mdn = 5.53$) being the fastest (No Boost: $M = 6.24, SD = 3.46, Mdn = 5.61$, Own Words: $M = 6.66, SD = 3.67, Mdn = 5.39$, Read All: $M = 7.77, SD = 4.87, Mdn = 6.67$).

Looking at educational background, the number of participants who correctly identified the article as fake news was relatively similar (high school: 65.3 \%, bachelor's degree: 60 \%, master's degree: 51.5 \%, doctorate: 55.6 \%, other forms of education: 60 \%).

Using Aslett’s scales, the sample shows a slightly liberal bias ($M = -0.41, SD = 2.18$) with 23 extreme conservatives, 36 extreme liberals, and 30 participants who did not report their ideology. Digital literacy scores of participants ranged from 8 to 44 ($M = 26.7, SD = 5.99$).

%\subsection{Boosts}

\subsection{Results}

The majority of the participants (59. 2 \%) were able to correctly identify the presented articles as fake news, while 31.3 \% believed the article to be true and the remaining 9.5 \% could not determine its veracity. This is a slightly higher percentage than was observed in Aslett's study 5, which may be partially explained by the boost interventions. However, we believe it is more likely to relate to the way in which news articles were sampled (see discussion below).

Table \ref{tab:boost_perf} shows the mean NewsGuard scores for the search results by condition. There appears to be a slight increase in %performance 
the score for the 'Multiple Queries' and 'Read All' conditions. However, the ANOVA results indicate that these differences are not statistically significant $(F(4, 1680) = 0.219, p = 0.931)$.

Similar trends are observed when examining the mean number of queries submitted by users. The 'No Boost' condition had the fewest queries overall, while the boost conditions saw slightly higher query counts. Again, the 'Read All' condition was the joint highest in terms of number of queries submitted. The percentage of participants who correctly identified the article also varied across conditions. The 'Read All' condition had the highest percentage, but the other three boost conditions were actually lower than the 'No Boost' condition. This suggests that the boosts did not have the effect we had expected, but advising users to read all of the search listings before crafting their queries holds the most promise. We did not conduct statistical tests on number of queries or task success because they were not included in our pre-registered analysis plan. This decision was made to avoid drawing misleading conclusions from multiple comparisons that were not planned in advance.

\begin{table}[ht]
\centering
\caption{Summary of Mean and SD of NewsGuard Scores, Num of Queries per User and Percentage of Participants who Identifed their Article as Fake News by Condition. Highest values are bolded}
\label{tab:boost_perf}
\resizebox{\columnwidth}{!}{%
\begin{tabular}{lccccc}
\toprule
\textbf{Cond.} & \textbf{Mean NG} & \textbf{SD} & \textbf{\parbox[c]{1.5cm}{\centering Mean\\ Queries}} & \textbf{SD} & \textbf{\% Ident.} \\
\midrule
No Boost           & 90.4  & 13.7 & 1.27  & 0.686 & 60.0\% \\
Read 1st           & 90.3  & 18.2 & \textbf{1.47}  & 0.910 & 57.1\% \\
Own Words          & 90.8  & 14.6 & 1.37  & 0.888 & 55.1\% \\
Mult. Queries   & 91.3  & 13.3 & 1.42  & 0.683 & 51.3\% \\
Read All           & \textbf{91.9}  & 15.8 & \textbf{1.47}  & 0.878 & \textbf{77.8\%} \\
\toprule
\end{tabular}}
\end{table}

Our setup also enabled us to explore behaviours not captured in the Aslett data, such as click-through data. Table \ref{tab:click_metrics} shows how click-based metrics differ across the experimental conditions. The first metric is the mean NewsGuard score for the viewed results, and the second is the percentage of clicked results that have an associated NewsGuard score. The second metric assumes that leading media outlets have available NewsGuard scores, and that unknown results are likely of lower quality. The findings align with those above: the 'No Boost' condition scores the lowest in both metrics, while 'Read All' performs well on both counts, but again the differences are not significant $F(4, 163) = 0.685, p = .603$.

\begin{table}[ht]
\centering
\caption{Summary of Mean and SD of NewsGuard Scores of the pages participants clicked on and percentage of clicked results with associated NewsGuard score. Highest values are bolded}
\label{tab:click_metrics}
\begin{tabular}{lccccc}
\toprule
\textbf{Cond.} & \textbf{Mean NG} & \textbf{SD} & \textbf{\% with Score} \\
\midrule
No Boost           & 86.2  & 15.9 & 67.7  \\
Read 1st           & 88.4  & 21.3 & 83.7  \\
Own Words          & \textbf{93.0}  & 9.0 & 69.8  \\
Mult. Queries      & 87.7  & 18.1 & 79.2  \\
Read All           & 90.7  & 16.6 & \textbf{84.4}   \\
\toprule
\end{tabular}
\end{table}

Finally, a correlation analysis reveals a negative relationship between the overall NewsGuard score (\texttt{avg\_score}) and the Jaccard overlap for headlines (\(-0.26\), \(p < 0.0001\)), while the correlations between \texttt{avg\_score} and the first paragraph (\(-0.09\), \(p = 0.16\)) and full text (\(-0.12\), \(p = 0.06\)) are weaker, suggesting that NewsGuard scores are more closely aligned with headline content than with other document parts.

\section{Discussion}

Taking the findings from the various investigations together reveals that validating misleading news articles is a challenging task. In our study fewer than 60\% %just over 40\% 
of participants were able to say with certainty that the article they were assigned was fake news and over 30\% believed it to be truthful. Aslett et al. discovered even higher percentages of participants believing misleading news.

The evidence indicates that searching to validate news is not per se problematic. The three studies (i.e., Sections \ref{sec:aslett}, \ref{sec:simulation} and \ref{sec:user_study}) consistently show that the way people create search queries impacts their ability to evaluate news articles. Using vocabulary from the source article, particularly the headline, often leads to lower-quality search results and increased belief in misinformation. Analysis of Aslett et al.'s data revealed that when participants believed a fake news article, their queries closely matched the headline. Although the headline+first paragraph conditions in the simulation performed best, further analysis revealed that higher quality results were linked to queries with less overlap with both the headline and initial paragraph (see Table \ref{tab:jaccard_scores}). Similar correlations were found in the boost study.

In general, the boost strategies used in this study showed limited success in modifying user behaviour and outcomes, a result that was unexpected given the effectiveness of similar interventions in influencing other search behaviours and outcomes \cite{bink2024balancing, ortloff2021effect}, including query generation \cite{harvey2015learning}. %The boost strategies used in this study showed limited success in modifying user behaviour and outcomes, a result that was unexpected given the effectiveness of similar interventions in influencing other search behaviours and outcomes \cite{bink2024balancing, ortloff2021effect}, including query generation \cite{harvey2015learning}. 
We consider reasons for this lack of impact. One potential explanation concerns how the messaging was perceived. While some studies suggest that boosts with tips with non obvious knowledge (e.g., informing users that content from trusted sources, such as .gov domains, is more reliable than .com \cite{ortloff2021effect, zimmerman2019privacy}) are acted on, it is possible that our boost messages were seen as unnecessary by certain participants. This may be attributed to overconfidence in their ability to identify misleading news, which is a well-documented phenomenon \cite{latham2008broken}. Participants who felt assured in their own information literacy skills might have dismissed the tips as irrelevant or patronising, thinking, ``I don't need this''. 

In light of this, interventions that promote intellectual humility—encouraging participants to acknowledge the limits of their knowledge—may be more effective. These could include boost interventions, similar to those proposed by Rieger et al.'s study \cite{rieger2021item}, or social nudges, where users compare their performance to that of experts, as seen in Bateman et al's work \cite{bateman2012search}.% We found that participants who were encouraged to read the returned search listings in full ('Read All') were more likely to generate better queries and seemed to some extent to achieve improved outcomes. %This suggests that interventions fostering openness to guidance, rather than reinforcing a sense of expertise, could enhance the effectiveness of such strategies.

Despite the lack of significant results, consistent trends in the data suggest that encouraging users to 'Read All' search results before forming a query is beneficial. Participants in this condition not only identified misleading articles most frequently but also submitted the most queries on average. Additionally, the articles they viewed were more likely to have an associated NewsGuard score, and the NewsGuard scores for the returned results were the highest in this condition.

It is important to note that 'Read All' was an intervention targeting query reformulation, and many participants submitted only a single query. Therefore, we suspect that for those influenced by the boost, it likely not only affected their querying behaviour but also implicitly encouraged them to interact with a broader set of results, promoting a more thorough approach to information evaluation. Combining this boost with one aimed at improving the initial query could potentially enhance its effectiveness.

\section{Limitations}

Although we present three complementary studies of different types, there are a number of limitations to our work, which we will discuss here along with their potential impact.

One limitation is that we focused exclusively on fake news. Our decision to centre the study on fake news was driven by the desire to build upon the main message from Aslett et al's work -that searching made people more likely to place their faith in misleading news. However, the querying strategies we explored may not be universally applicable to all types of articles. For instance, using vocabulary directly from the source article in queries may be particularly problematic for fake news but may have the opposite effect if the article is trustworthy.

In the simulated study, we used the same articles as those in Aslett and colleague’s work. Although these articles were outdated, we chose them to ensure comparability with previous studies. In hindsight, it might have been better to use the approach we applied in our final user study, which involved more current articles. Even with this method, the articles were not entirely ``fresh'', as the fact-checking process meant the articles were already 1-3 days old by the time the study took place. This is comparable to the original Aslett study.

Despite these limitations, the patterns we observed were consistent across all three analyses. Encouraging users to use vocabulary not directly contained in the source article appears to be key to improving the quality of the search results they receive.

Another limitation is that our user study focused on only four articles. While we aimed to capture differences in user behaviour by collecting multiple data points per article, the power analysis indicated that 200 data points were needed, with 50 per article being plausible. However, these four articles may not be representative of all fake news articles, and future research could expand the study to include a broader sample. In fact, we believe our article sampling process may have been biased towards selecting more obviously misleading articles, as we worked to achieve high agreement between three researchers and professional fact-checkers. This process proved challenging, as many article authors crafted misleading narratives without making explicit claims. Instead, they carefully curated quotes to advance a particular narrative. We selected articles where the claims were more clearly defined, which makes it all the more concerning that a significant percentage of our participants still rated these articles as trustworthy.

Additionally, we only simulated query generation based on specific parts of the document. While this was a sensible first step, there are many other strategies that could be explored in future studies, such as crafting queries to specifically reflect claims or involving negation.

We tested only four boost strategies.
%, none of which proved to be significantly effective, although there were some indications that they positively influenced behaviour. 
Other approaches, not explored in this study but mentioned in our discussion, might offer more promising results and could be valuable for future research.

Lastly, we used NewsGuard as a proxy for search quality, which we believe was a well-justified choice. However, NewsGuard does not capture the semantic aspect of search quality, such as whether queries drift off topic. We are currently developing other metrics that focus on the claims made in search results and how they confirm or contradict those in the source document. We believe these new metrics will provide deeper insights into user behaviour.

Overall, while our studies provide valuable insights, these limitations highlight areas for improvement and further exploration in future work.

\section{Ethical Considerations}

The choice of boost interventions as a framework to assist users in making better decisions leverages behavioural patterns without restricting the searcher's freedom of choice. We consider this a key strength when compared to alternative approaches such as the filtering of results.

While boosts are typically associated with transparency and providing knowledge and competences, they can still negatively impact on users. This is demonstrated in Table \ref{tab:boost_perf} which shows that in three of the four boost conditions participants were less likely to identify fake news.

\section{Future work and Conclusions}

Beyond testing the ideas presented in the discussion, an obvious direction for future research is to explore how Generative AI systems, which are increasingly central to information-seeking, fit into this ecosystem. AI agents, such as co-pilots, could potentially help users create better queries, especially if they understand the task at hand. Furthermore, generative AI interfaces like ChatGPT may be used instead of traditional search engines to validate news articles. This raises many questions, such as how users would interact with these systems for this purpose and how successful they would be.

In summary, our work has examined the influence of querying behaviour on people's efforts to validate news articles. The evidence strongly suggests that the way queries are crafted plays a role, and there is potential for systems to be designed to help users improve this process. While  the best way to achieve this is not yet clear, it is clear to us that this approach could complement other measures, such as providing media literacy training, promoting fact-checking tools, integrating source credibility indicators, encouraging scepticism through nudges, and fostering collaboration with expert networks, to empower users and enhance their information literacy.

\section{Open Science}
We have made all resources including the data, code, articles and processes available in two anonymised github repositories:\\

See \url{https://github.com/markusbink/sigir-aslett-misinfo/} for Sections \ref{sec:aslett} and \ref{sec:user_study}.\\
See \url{https://github.com/MarcosFP97/sim-sigir} for Section \ref{sec:simulation}.

\newpage

\bibliographystyle{ACM-Reference-Format}
\bibliography{bibliography}

\end{document}